# Horizontal gene transfer may explain variation in $\theta_s$

Arising from Martincorena, I., Seshasayee, A. S. N. & Luscombe, N. M. *Nature* **485**, 95-98 (2012)

Rohan Maddamsetti[1,2], Philip J. Hatcher[3], Stéphane Cruveiller[4], Claudine Médigue[4], Jeffrey E. Barrick[2,5] & Richard E. Lenski[1,2]

[1]Ecology, Evolutionary Biology, and Behavior Program, Michigan State University, East Lansing, Michigan 48824, USA. [2]BEACON Center for the Study of Evolution in Action, Michigan State University, East Lansing, Michigan 48824, USA. [3]Department of Computer Science, University of New Hampshire, Durham, NH 03824, USA. [4]CNRS-UMR 8030 and Commissariat à l'Energie Atomique CEA/DSV/IG/Genoscope LABGeM, 91057 Evry Cedex, France. [5]Institute for Cellular and Molecular Biology, Department of Chemistry and Biochemistry, The University of Texas at Austin, Austin, TX 78712, USA.

Martincorena *et al.*[1] estimated synonymous diversity ($\theta_s = 2N\mu$) across 2,930 orthologous gene alignments from 34 *Escherichia coli* genomes, and found substantial variation among genes in the density of synonymous polymorphisms. They argue that this pattern reflects variation in the mutation rate per nucleotide ($\mu$) among genes. However, the effective population size (N) is not necessarily constant across the genome[2]. In particular, different genes may have different histories of horizontal gene transfer (HGT), whereas Martincorena *et al*. used a model with random recombination[3] to calculate $\theta_s$. They did filter alignments in an effort to minimize the effects of HGT, but we doubt that any procedure can completely eliminate HGT among closely related genomes, such as *E. coli* living in the complex gut community.

    Here we show that there is no significant variation among genes in rates of synonymous substitutions in a long-term evolution experiment with *E. coli* and that the per-gene rates are not correlated with $\theta_s$ estimates from genome comparisons. However, there is a significant association between $\theta_s$ and HGT events. Together, these findings imply that $\theta_s$ variation reflects different histories of HGT, not local optimization of mutation rates to reduce the risk of deleterious mutations as proposed by Martincorena *et al*.

    We replicated the $\theta_s$ estimates of Martincorena *et al.* on a set of 2,853 single-copy orthologous gene alignments[4] from 10 *E. coli* genomes. We then identified all synonymous substitutions in these genes in clones isolated from 12 independently evolved populations after 40,000 generations of a long-term experiment[5-7]. Six clones derived from lineages that evolved mutations in *mutS*, *mutL*, or *mutT*[6,7], and these "mutators" accounted for 1,055 of the 1,069 synonymous substitutions that we identified. We compared the observed cumulative probability distribution of synonymous substitutions across this gene set with the distribution expected under the null hypothesis of a uniform point-mutation rate (Fig. 1), and that difference is non-significant (Kolmogorov-Smirnov test, $D = 0.0281$, $P = 0.21$).

    In contrast, the difference between the observed distribution and that expected under the alternative hypothesis that $\mu \propto \theta_s$ is extremely significant (Fig. 1; Kolmogorov-Smirnov test, $D = 0.244$, $P < 10^{-15}$). Importantly, this difference holds when data for the

four mismatch-repair, *mutS* and *mutL*, and two *mutT* mutators are analyzed separately ($P < 10^{-15}$ and $P < 10^{-8}$, respectively). Hence, rejection of this hypothesis is independent of the particular mutational signatures of each mutator class.

We then looked for evidence of HGT in *Escherichia* and *Shigella* orthologs of these genes using a dataset on HGT in the human microbiome[8]. We assigned these events to genes in the ancestral genome of the evolution experiment[9] on the basis of sequence homology (BLAST E-value $< 10^{-10}$). The $\theta_s$ values for the 11 genes with HGT events were significantly higher than for the other 2,842 genes (one-tailed Mann-Whitney test, $P = 0.022$) (Fig. 1).

Martincorena *et al.* also showed that genes with low $\theta_s$ have low values of Tajima's D and tend to be expressed at higher levels, characteristics expected for housekeeping genes subject to strong purifying selection. These observations are also consistent with HGT, because such genes will resist the influx of alleles more effectively than genes that face weak or variable selection pressures. In summary, our analyses offer no support for the hypothesis that point-mutation rates vary among genes as postulated by Martincorena *et al.* Instead, the observed variation between genes in synonymous substitutions may be explained by their different histories of HGT.

**Figure 1 | Synonymous substitution rates reflect uniform mutation rate across genes in lab-evolved clones and gene-transfer events in natural isolates.** Each observed or hypothetical series shows the cumulative proportion of synonymous substitutions within increasing subsets of 2,853 *E. coli* genes sorted and ranked by their $\theta_s$ values. The filled (and overlapping) symbols show the observed distribution of synonymous mutations in 12 independently evolved *E. coli* genomes after 40,000 generations; those genes with evidence for HGT events in natural isolates are black and others are green. The dashed line shows the null hypothesis of a uniform point-mutation rate; the dashed and dotted curve shows the alternative hypothesis where the point-mutation rate is proportional to $\theta_s$.

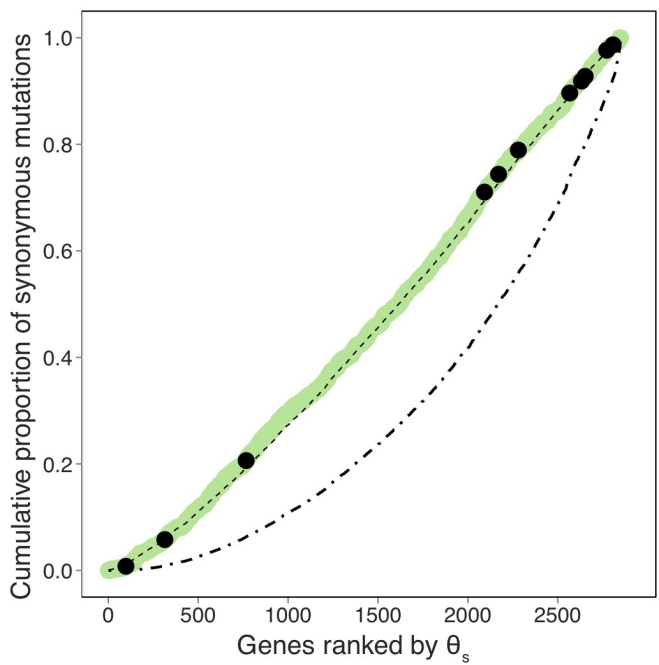